\documentclass[aps,prl,twocolumn,preprintnumbers,amsmath,amssymb,nofootinbib,superscriptaddress,notitlepage,10pt]{revtex4-1}

\usepackage[utf8]{inputenc}
\usepackage[sort&compress]{natbib}
\usepackage{comment}
\usepackage{amssymb}
\usepackage{color}
\usepackage{rotating}
\usepackage{amsmath}
\usepackage{ulem}
\usepackage{overpic}
\usepackage{graphicx}
\usepackage{epstopdf}
\usepackage{dcolumn}
\usepackage{bm}
\usepackage{chngpage}
\usepackage{multirow}
\usepackage{slashed}
\usepackage{indentfirst}
\usepackage{booktabs}
\usepackage{amssymb,amsfonts}
\usepackage{amsmath}
\usepackage{color}
\usepackage{hyperref}
\usepackage{txfonts}
\usepackage{epsfig}
\usepackage{slashed}
\usepackage{color}
\usepackage{appendix}
\usepackage{graphicx}

\begin{document}

\title{Discovery of the doubly charmed $T_{cc}^+$ state implies a triply charmed  $H_{ccc}$ hexaquark state}

\author{Tian-Wei Wu}
\affiliation{School of Physics, Beihang University, Beijing 102206, China}
\author{Ya-Wen Pan}
\affiliation{School of Physics, Beihang University, Beijing 102206, China}

\author{Ming-Zhu Liu}
\affiliation{School of Space and Environment, Beihang University, Beijing 102206, China}
\affiliation{School of Physics, Beihang University, Beijing 102206, China}

\author{Si-Qiang Luo}
\affiliation{School of Physical Science and Technology, Lanzhou University, Lanzhou 730000, China}

\author{Li-Sheng Geng}\email{lisheng.geng@buaa.edu.cn}
\affiliation{School of Physics, Beihang University, Beijing 102206, China.}
\affiliation{Beijing Key Laboratory of Advanced Nuclear Materials and Physics, Beihang University, Beijing, 102206, China}
\affiliation{School of Physics and Microelectronics, Zhengzhou University, Zhengzhou, Henan 450001, China}

\author{Xiang Liu}\email{xiangliu@lzu.edu.cn}
\affiliation{School of Physical Science and Technology, Lanzhou University, Lanzhou 730000, China}
\affiliation{Research Center for Hadron and CSR Physics, Lanzhou University and Institute of Modern Physics of CAS, Lanzhou 730000, China}
\affiliation{Lanzhou Center for Theoretical Physics, Lanzhou University, Lanzhou, Gansu 730000, China}

\date{\today}


\begin{abstract}
  \rule{0ex}{3ex}
  The  doubly charmed exotic state $T_{cc}$ recently discovered  by the LHCb Collaboration could well be a $DD^{*}$ molecular state long predicted in various theoretical models, in particular, the $DD^*$ isoscalar axial vector molecular state predicted in the one-boson-exchange model. In this work, we study the $DDD^*$  system in the Gaussian Expansion Method with the $DD^*$ interaction derived from the one-boson-exchange model and  constrained by the  precise binding energy of  $273\pm63$ keV of $T_{cc}$ with respect to the $D^{*+}D^0$ threshold. We  show the existence of a $DDD^*$ state with a binding energy of a few hundred keV and spin-parity $1^-$. Its main decay modes are $DDD\pi$ and $DDD\gamma$. The existence of such a state could in principle be confirmed with the upcoming LHC data and will unambiguously determine the nature of the $T_{cc}^+$ state and of the many exotic state of similar kind, thus deepening our understanding of the non-perturbative strong interaction.
\end{abstract}

\maketitle

{\it Introduction.---}
Starting from the discovery of  $D_{s0}^*(2317)$~\cite{Aubert:2003fg} and $X(3872)$~\cite{Choi:2003ue} in 2003, a large number of the so-called exotic states that do not fit into the conventional quark model have been observed, which have led to intensive studies both theoretically and experimentally~\cite{Brambilla:2019esw,Liu:2013waa,Hosaka:2016pey,Chen:2016qju,Guo:2017jvc,Liu:2019zoy}. The latest addition to this long list  is the $T_{cc}^+$ state reported by the LHCb Collaboration  at the
European Physical Society conference on high energy physics
2021~\cite{LHCb1,LHCb2}. This state has a minimum quark content of $cc\bar{u}\bar{d}$ with a binding energy of $B=273\pm 61\pm  5^{+11}_{-14}$ keV with respect to the $D^{*+}D^0$ threshold and a decay width of $\Gamma=410\pm 165\pm  43^{+18}_{-38}$ keV. Although such a doubly charmed state has long been anticipated theoretically~\cite{Semay:1994ht,Janc:2004qn,Vijande:2007rf,Lee:2009rt,Yang:2009zzp,Li:2012ss,Karliner:2017qjm,Wang:2017uld,Junnarkar:2018twb,Liu:2019stu}, it has remained elusive experimentally until now. Being the first doubly charmed tetraquark state, its discovery will undoubtedly usher in a new era in hadron spectroscopy studies and   advance our understanding of the non-perturbative strong interaction. 

The measured binding energy and preferred quantum numbers of the $T_{cc}$ state are in very good agreement with our predictions based on the one-boson-exchange(OBE) model~\cite{Li:2012ss,Liu:2019stu}, thus qualifies as a $DD^*$ molecule with $I(J^P)=0(1^+)$. An urgent question of high relevance is to understand the nature of this state, how to distinguish the various interpretations, and study the consequences.

\begin{figure}[htpb]
  \centering
  \includegraphics[scale=0.1]{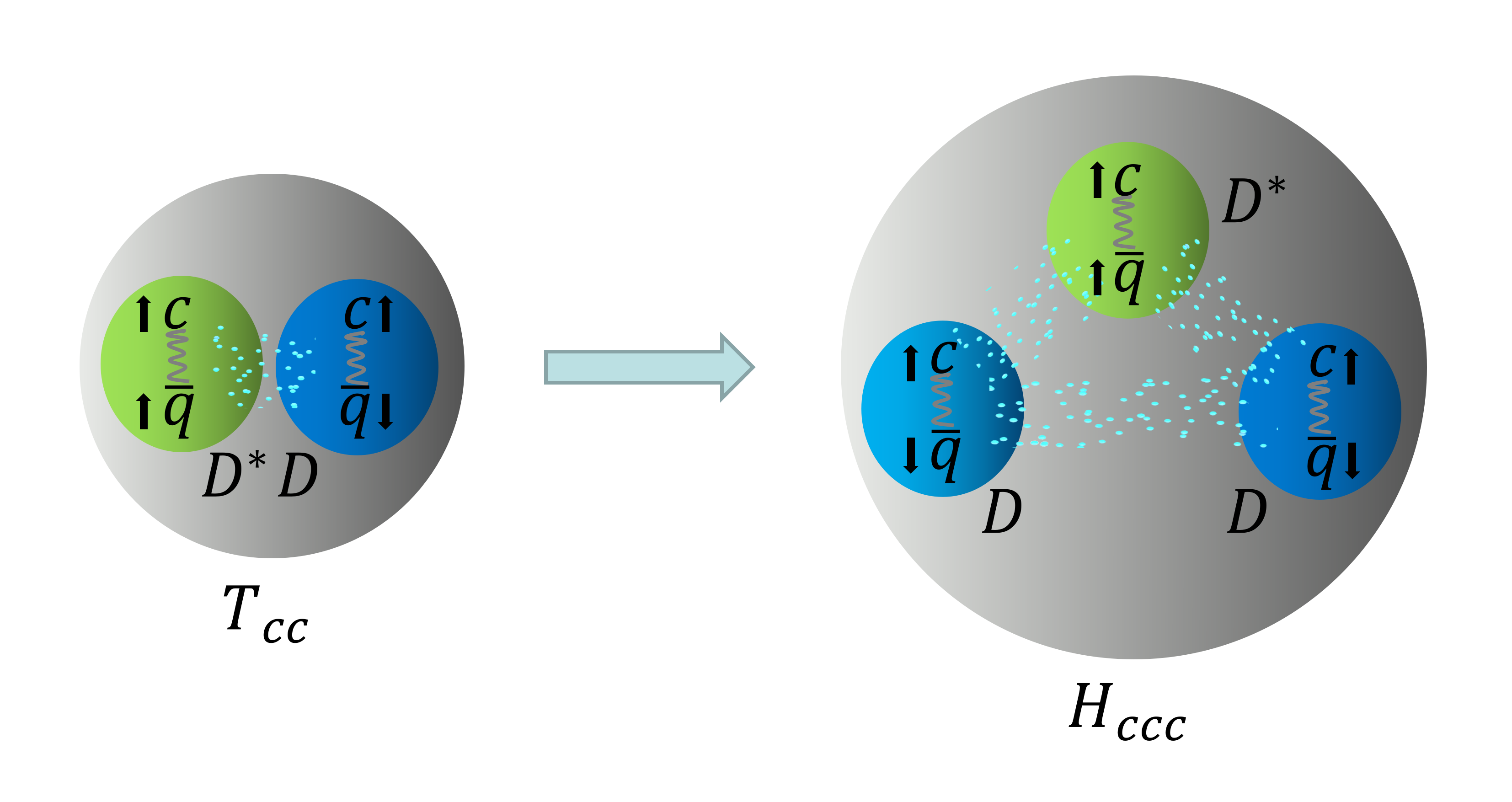}
  \caption{From $T_{cc}$ (as a $DD^*$ molecule) to $H_{ccc}$ (as a $DDD^*$ molecule).}
  \label{two2three}
\end{figure}
Being close to some certain two-hadron thresholds is only a necessary but not sufficient condition for a particular hadron to be of molecular nature. Taking  $X(3872)$ as one example, after almost 20 years of extensive studies, there is still ongoing heated debate about its true nature being either  a conventional $c\bar{c}$ charmonium, a $\bar{D}D^*$ molecule, a compact tetraquark state, or a combination of them. In a series of recent studies~\cite{SanchezSanchez:2017xtl,Ren:2018pcd,MartinezTorres:2018zbl,Wu:2019vsy,Huang:2019qmw,Wu:2020rdg,Pang:2020pkl,Wu:2020job,Wu:2021ljz,Wu:2021gyn}, we argued that one way to check the molecular nature of certain exotic hadrons is to search for existence of multi-hadron molecules built from the same constituents, in the way that atomic nuclei are bound states of multi-nucleons~\footnote{For a concrete demonstration that one can confidently deduce the existence of triton from that of deuteron using either the OBE model or a phenomenological model to describe the nucleon-nucleon interaction, see Ref.~\cite{Wu:2021dwy}.}. More specifically, it was shown that if $D_{s0}(2317)$ is dominantly a $DK$ bound state, then $DDK$, $D\bar{D}K$, and $D\bar{D}^*K$ states should exist~\cite{MartinezTorres:2018zbl,Wu:2019vsy,Wu:2020job}~\footnote{The predicted $DDK$ state has a minimum quark content of $cc\bar{s}\bar{u}/\bar{d}$, isospin 1/2, and spin-parity $0^-$. Such a state has recently been searched for by the Belle Collaboration~\cite{Belle:2020xca}.}. Similarly, if the latest $T_{cc}^+$ state is indeed a $DD^*$ molecule, then it is very likely that  a $DDD^*$  bound state exists (see Fig.~\ref{two2three}). Given the capacity of the LHCb experiment, such a state could very well be discovered in the near future and thus not only provide a highly nontrivial check on the molecular nature of the $T_{cc}^+$ state but also deepen our understanding of the strong interaction.

In this work, with  the latest experimental measurements~\cite{LHCb1,LHCb2}, we  fix the $DD^*$ interaction provided by the time-honored OBE model, and study the $DDD^*$ system using the Gaussian Expansion Method. 

{\it Theoretical formalism.---}
 The Gaussian Expansion Method has been widely used to solve three-, four- and even five-body  problems, because of  its high precision and rapid convergence~\cite{Hiyama:2010zzd}. In this framework, the three-body $DDD^*$ system is described by the following Schr\"{o}dinger equation
\begin{equation}
\hat{H}\Psi=E\Psi,
\end{equation}
where the Hamiltonian $\hat{H}$ includes the kinetic term and three two-body interaction terms
\begin{equation}
  \hat{H}=T+V_{DD}+V_{DD^*}+V_{DD^*}.
\end{equation}
In order to solve the Schr\"{o}dinger equation, we have to first specify the two-body interactions.
\begin{table}[!h]
\caption{Couplings of the light mesons of the OBE model
  ($\pi$, $\sigma$, $\rho$, $\omega$) to the heavy $D/D^*$ mesons.
  For the magnetic-type coupling of the $\rho$ and $\omega$ vector mesons
  we have used the decomposition
  $f_{\rho (\omega)} = \kappa_{\rho (\omega)}\,g_{\rho (\omega)}$.
  $M$ (in units of MeV) refers to the mass scale involved in the magnetic-type couplings~\cite{Liu:2019stu}.
}
\begin{tabular}{c|c}
  \hline \hline
  Coupling  & Value for $D$/$D^*$ \\
  \hline
  $g$ & 0.60 \\
  $g_{\sigma}$ & 3.4 \\
  $g_{\rho}$ & 2.6 \\
  $g_{\omega}$ & 2.6 \\
  $\kappa_{\rho}$ & 4.5 \\
  $\kappa_{\omega}$ & 4.5 \\
  $M$ & 1867 \\
  \hline \hline
\end{tabular}
\label{tab:couplings}
\end{table}
In our present work, both the $DD$ interaction and the $DD^*$ interaction are derived from the OBE model. In Ref.~\cite{Wu:2019vsy}, the $DD$ OBE potential has been derived with the exchange of $\sigma$, $\rho$, and $\omega$ mesons. 
For the $DD^*$ interaction, one can also exchange a $\pi$ meson in addition to  the $\sigma$, $\rho$ and $\omega$ exchanges~\cite{Liu:2019stu}. It should be noted that the $DD^*$ interaction of Ref.~\cite{Liu:2019stu} generates   a molecular $DD^*$ state with a cutoff of 1.01 GeV, which was fixed by reproducing the binding energy 4.0 MeV of $X(3872)$ with respect to the $D\bar{D}^{*}$ threshold. A detailed description of the OBE potential used can be found in Refs.~\cite{Wu:2019vsy,Liu:2019stu}.  With the relevant couplings between $DD^{(*)}$ and the exchanged mesons fixed (as shown in Table \ref{tab:couplings}), the only free parameter is the cutoff related to the regulator function needed to take into account the finite size of exchanged mesons. More specifically, we use a regulator function of the following form
\begin{eqnarray}
  F(q, m, \Lambda)=
  {\left( \frac{\Lambda^{2}-m^2}{\Lambda^{2}-{q}^2} \right)}
  \label{Eq:FF} 
\end{eqnarray}
where $m$ is the mass of the exchanged meson  (see Table \ref{tab:exmasses}) and $\Lambda$ the cutoff.

\begin{table}[!h]
\caption{Masses and quantum numbers of the light mesons
  of the OBE model ($\pi$, $\sigma$, $\rho$, $\omega$)
  and the heavy mesons $D$ and $D^*$~\cite{ParticleDataGroup:2020ssz}.}
\begin{tabular}{c|cc}
  \hline\hline
  Light Meson  & $I^{G}\,(J^{PC})$  & M (MeV) \\
  \hline
  $\pi$ & $1^{-}$ $({0}^{-+})$ & 138 \\
  $\sigma$ & $0^{+}$ $({0}^{++})$ & 600 \\
  $\rho$ & $1^{+}$ $({1}^{--})$ & 770 \\
  $\omega$ & $0^{-}$ $({1}^{--})$ & 780 \\
  \hline 
  Heavy Meson & $I (J^P)$ & M (MeV) \\
  \hline
  $D$ & $\frac{1}{2}(0^-)$ & 1867.24 \\
  $D^*$ & $\frac{1}{2}(1^-)$ & 2008.56 \\
  \hline \hline
\end{tabular}
\label{tab:exmasses}
\end{table}
First, we slight fine-tune the cutoff (from the value of 1.01 GeV fixed by reproducing a binding energy of 4 MeV for the $D\bar{D}^*$ bound state assigned to be $X(3872)$~\cite{Liu:2019stu}) taking advantage of  the latest experimental data~\cite{LHCb1,LHCb2}. As the $T_{cc}$ state is found about 0.3 MeV below the $D^{*+}D^0$ threshold and the difference between the thresholds of $D^+ D^{*0}$ and $D^{*+}D^0$ is 1.41 MeV, we study three different binding energy scenarios for the $DD^*$ binding energy, i.e., 0.3 MeV, 1.0 MeV, and 1.7 MeV. The so-determined cutoffs for these three scenarios considering only $S$-wave interactions and $S-D$ mixings are given in Table~\ref{tab:cutoff}. In the same table, we also provide the corresponding root-mean-square (RMS) radius of the $T_{cc}^+$ state. Two things are noteworthy. First, the RMS radius ranges from 3 to 6 fm, consistent with the expectation for a  molecular state whose size should be larger than the sum of its constituents. Second, the impact of $S-D$ mixing is small at the two-body level, consistent with the analysis of Ref.~\cite{Li:2021zbw}. Based on this observation, we only consider $S$-wave interactions among the $D^{(*)}$ mesons in the following study of the $DDD^*$ system.

\begin{table}[!h]
\caption{Cutoffs for three different binding energy scenarios and with/without considering $S-D$ mixing. The binding energies ($B$) are in units of MeV and RMS radii $r$ in units of fm. 
}\label{tab:cutoff}
\begin{tabular}{cccccc}
  \hline\hline
  $\Lambda$(Only $S$)  & $B$  & $r_{DD^*}$ &$\Lambda$($S-D$)  & $B$  & $r_{DD^*}$ \\
  \hline
    976 & 0.3&5.94&945 & 0.3&6.09 \\
     998 & 1.0&3.47 &  970 & 1.0&3.55  \\
    1013 & 1.7&2.72 & 986 & 1.7&2.81 \\
  \hline \hline
\end{tabular}

\label{tab:masses}
\end{table}

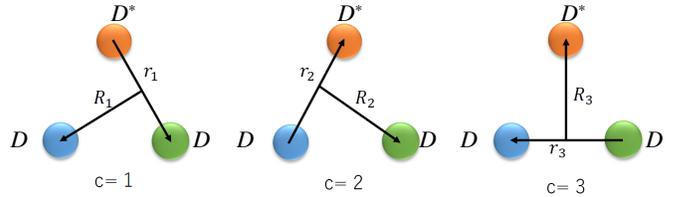
\begin{figure}[!h]
  \centering
  \begin{overpic}[scale=0.35]{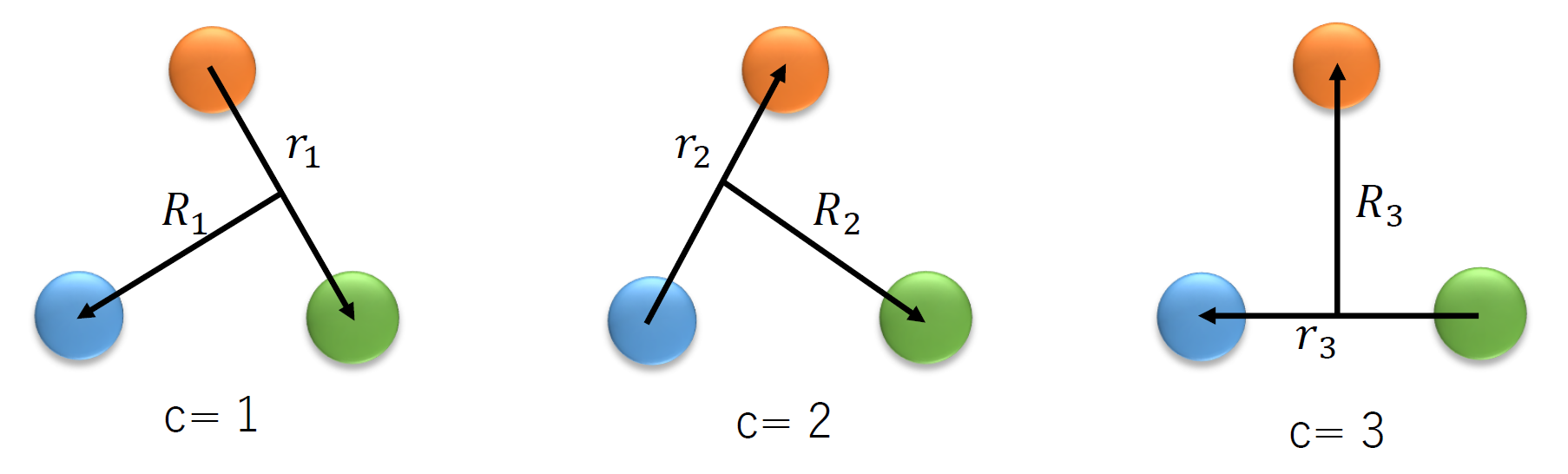}
  		\put(-3,9){$D$}
  		\put(26,9){$D$}
  		\put(13,29){$D^*$}
		\put(33,9){$D$}
		\put(62,9){$D$}
		\put(49,29){$D^*$}
		\put(68.5,9){$D$}
		\put(98,9){$D$}
		\put(84.5,29){$D^*$}
  \end{overpic}
  \caption{Three permutations of the Jacobi coordinates for the $DDD^*$ system.}
  \label{Jac}
\end{figure}

As all the two-body interactions have been specified, we employ the GEM to solve the Schr\"odinger equation. The three-body wave functions can be constructed in Jacobi coordinates as 
\begin{equation}
    \Psi=\sum_{c=1}^{3}\Psi(\bf{r}_c,\bf{R}_c),
\end{equation}
where $c=1-3$ is the label of the Jacobi channels shown in Fig.~\ref{Jac}. In each Jacobi channel the wave function $\Psi(\mathbf{r}_c,\mathbf{R}_c)$ reads
\begin{equation}
    \Psi(\mathbf{r}_c,\mathbf{R}_c)=C_{c,\alpha}H^c_{t,T}\Phi_{lL,\lambda}(\mathbf{r}_c,\mathbf{R}_c)
\end{equation}
where $C_{c,\alpha}$ is the expansion coefficient and the $\alpha=\{nN,tT,lL\lambda\}$ labels the basis number with the configuration sets of the Jacobi channels. $H^c_{t,T}$ is the  three-body isospin wave function where $t$ is the isospin of the subsystem in Jacobi channel $c$ and $T$ is the total isospin. 

The three-body spatial wave function $\Phi(\mathbf{r}_c,\mathbf{R}_c)$ is constructed by two two-body wave functions as
\begin{equation}
\begin{split}
       \Phi_{lL,\lambda}(\mathbf{r}_c,\mathbf{R}_c)&=[\phi_{n_cl_c}^{G}(\mathbf{r}_c)\psi_{N_cL_c}^{G}(\mathbf{R}_c)]_{\lambda},\\
     \phi_{nlm}^{G}(\mathbf{r}_c)&=N_{nl}r_c^le^{-\nu_n r_c^2} Y_{lm}({\hat{r}}_c),\\
     \psi_{NLM}^{G}(\mathbf{R}_c)&=N_{NL}R_c^Le^{-\lambda_n R_c^2} Y_{LM}({\hat{R}}_c).
\end{split}
\end{equation}
Here $N_{nl}(N_{NL})$ is the normalization constant of the Gaussian basis, $n(N)$  is the number of Gaussian basis used, $l(L)$ is the orbital angular momentum corresponding to the Jacobi coordinates $r (R)$, and $\lambda$ is the total orbital angular momentum.  

With the constructed wave functions, the Schr\"odinger equation can be transformed into a generalized matrix eigenvalue problem with the Gaussian basis functions
\begin{equation}\label{eigenvalue problem}
  [T_{\alpha \alpha'}^{ab}+V_{{\alpha \alpha'}}^{ab}-EN_{\alpha \alpha'}^{ab}]\,
  C_{b,\alpha'} = 0
  \, ,
\end{equation}
where $T_{\alpha \alpha'}^{ab}$ is the matrix element of kinetic energy, $V_{\alpha \alpha'}^{ab}$ is the  matrix element of potential energy, and $N_{\alpha \alpha'}^{ab}$ is the normalization matrix element.

{\it Results and Discussions.---}Considering only $S$-wave interactions, the corresponding configurations of the three Jacobi channels are given in Table \ref{tab:configs}. With these configurations and the OBE potentials specified above, we solve the Schr\"odinger equation in the GEM method and obtain the results shown in Table~\ref{tab:results}.  It is interesting to note that for all the three scenarios studied, the $DDD^*$ system is bound. Compared to the $DD^*$ system, the addition of a second $D$ meson  only increases the binding energy by about 23\%, 29\%, and 34\%, reflecting the fact that the $DD$ interaction is less attractive than the $DD^*$ interaction. This is corroborated by the observation that for all the three scenarios $r_{DD}$ is larger than $r_{DD^*}$ and $|\langle V_{DD^*}\rangle|$ is much larger than $|\langle V_{DD}\rangle|$.
\begin{table}[h!]
    \centering
        \caption{Quantum numbers of different Jacobi coordinate channels ($c = 1-3$) of the $DDD^*$ $I(J^{P})=\frac{1}{2}(1^{-})$ state, considering only $S$-wave interactions.\label{tab:configs}}
    \begin{tabular}{cccccccccc}
    \hline\hline\label{tab:configs}
     $c$ & $l$ & $L$ & $\Lambda$ & $t$ & $T$ & $J$ & $P$ & $n_{max}$ & $N_{max}$ \\
     1 & 0 & 0 & 0 & 0 & 1/2 & 1 & $-$ & 10 & 10\\
     1 & 0 & 0 & 0 & 1 & 1/2 & 1 & $-$ & 10 & 10\\
    2 & 0 & 0 & 0 & 0 & 1/2 & 1 & $-$ & 10 & 10\\
     2 & 0 & 0 & 0 & 1 & 1/2 & 1 & $-$ & 10 & 10\\
     3 & 0 & 0 & 0 & 1 & 1/2 & 1 & $-$ & 10 & 10\\
    \hline\hline
    \end{tabular}

\end{table}

\begin{table}[h!]
    \centering
    \caption{Binding energies, RMS radii and Hamiltonian expectation values of the  $DDD^*$ system with $I(J^{P})=\frac{1}{2}(1^{-})$ and S-wave OBE interactions.}\label{tab:results}
    \begin{tabular}{ccccccc}
    \hline\hline
          $\Lambda$(MeV) & $B$(MeV) & $r_{DD^{\ast}}$ & $r_{DD}$ & $\left \langle T \right \rangle$ & $\left \langle V_{DD^{\ast}} \right \rangle$ & $\left \langle V_{DD} \right \rangle$ \\ \hline
          976 & 0.37 & 8.20 & 10.57 & 10.32 & $-10.53$ & $-0.17$ \\
          998 & 1.29 & 5.11 & 6.72 & 20.48 & $-21.17$ & $-0.60$ \\
          1013 & 2.27 & 3.87 & 5.06 & 27.81 & $-29.11$ & $-0.98$ \\ \hline\hline
    \end{tabular}\\
\end{table}

\begin{table}[h!]
    \centering
    \caption{Binding energies, RMS radii and Hamiltonian expectation values of the doubly charged $I(J^P)=\frac{1}{2}(1^-)$ $DDD^*$ state  with $S$-wave OBE  and Coulomb interactions.}\label{tab:results2}
    \begin{tabular}{ccccccc}
    \hline\hline
          $\Lambda$(MeV) & $B$(MeV) & $r_{DD^{\ast}}$ & $r_{DD}$ & $\left \langle T \right \rangle$ & $\left \langle V_{DD^{\ast}} \right \rangle$ & $\left \langle V_{DD} \right \rangle$ \\ \hline
          976 & 0.15 & 8.78 & 10.41 & 6.83 & $-6.99$ & $0.02$ \\
          998 & 0.80& 7.88 & 10.60 & 15.10 & $-15.87$ & $-0.04$ \\
          1013 & 1.61 & 5.65 & 7.64 & 23.40 & $-24.72$ & $-0.28$ \\ \hline\hline
    \end{tabular}\\
\end{table}

Since the isospin of the studied $DDD^*$ system is 1/2, this system consists of two charged states, i.e.,
\begin{eqnarray}
   I(1/2,1/2)&:& \sqrt{\frac{2}{3}}D^+D^+D^{*0}-\sqrt{\frac{1}{6}}(D^+D^0+D^0D^+)D^{*+},\nonumber\\
   I(1/2,-1/2)&:&-\sqrt{\frac{2}{3}}D^0D^0D^{*+}+ \sqrt{\frac{1}{6}}(D^+D^{0}+D^0D^{+})D^{*0}.\nonumber
\end{eqnarray}
The Coulomb interaction may play a role for the doubly charged state which corresponds to the $I_3=1/2$ component. We include the Coulomb interaction for this state and find that the binding energies are 0.15, 0.80, and 1.61 MeV corresponding to the cutoff 0.967, 0.998, and 1.013 GeV, respectively. The results are shown in Table \ref{tab:results2}. The coulomb interaction makes the binding energy of the doubly charged state slightly smaller compared to the singly charged one, but is not strong enough to break it up. The main reason is that the $DD$ pair is widely separated at a distance of about 10 fm..

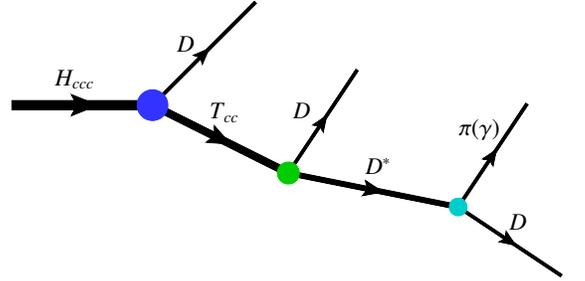
\begin{figure}[htpb]
\begin{center}
\begin{tabular}{cc}
\begin{minipage}[t]{0.8\linewidth}
\begin{center}
\begin{overpic}[scale=.8]{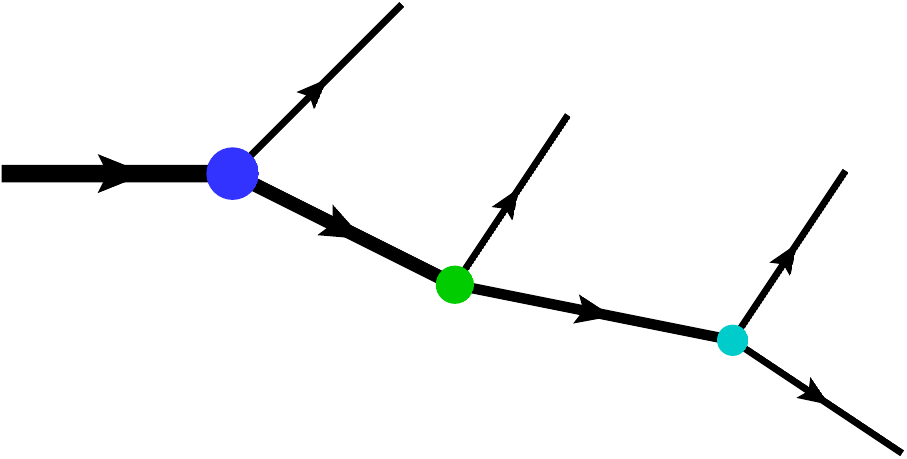}
		\put(90,9){$D$}
		
		\put(51,29){$D$}
    	\put(30,41){$D$}
		\put(36,29){$T_{cc}$}
		
		\put(8,35){$H_{ccc}$ }
		\put(81,26){$\pi(\gamma)$} \put(64,19){$D^{\ast}$}
\end{overpic}
\end{center}
\end{minipage}
\end{tabular}
\caption{ Decay mechanism of $H_{ccc}$    }
\label{decay}
\end{center}
\end{figure}

In principle the predicted triply charmed $H_{ccc}$ state can decay into $DDD\pi$  because the $DDD^*$ system could be viewed as a weakly bound $DT_{cc}$ state, in which the $T_{cc}$ state decays into $DD\pi$ as observed by the LHCb Collaboration~\cite{LHCb1,LHCb2}. Such a process is schematically shown in Fig.~\ref{decay}. Theoretically, as  $D^*$ can also decay into $D\gamma$, the $H_{ccc}$ state can also be observed in the $DDD\gamma$ mode. According to the LHCb measurements, the estimated  yield of $T_{cc}\to DD\pi$ with respect to that of $X(3872)\to D\bar{D}\pi$ is abut 1/20~\cite{LHCb2}. Naively the yield of $H_{ccc}\to DDD\pi$ with respect to that of $T_{cc}\to DD\pi$ might only be one order of magnitude smaller, thus accessible to future LHCb experiments.

{\it Summary and outlook.---}
The recently discovered doubly charmed $T_{cc}$ state is consistent with a $DD^*$ molecule predicted in the OBE model. The precisely measured binding energy with respect to the $D^{*+}D^0$ threshold allows one to fix the $DD^*$ interaction. We utilized this valuable information and studied the three-body $DDD^*$ system with the Gaussian Expansion Method. Our studies showed that the $DDD^*$ system is bound even taking into account the Coulomb interaction. We discussed the possible decay modes where the $DDD^*$ states can be discovered. We strongly encourage that this state be directly searched for at present and future experiments.

{\it Acknowledgement.---}
This work is partly supported by the National Natural Science Foundation of China under Grants No.11735003, No.11975041, No.11961141004,  and the fundamental Research Funds for the Central Universities.  XL is supported by the China National Funds for
Distinguished Young Scientists under Grant No. 11825503,
National Key Research and Development Program of China
under Contract No. 2020YFA0406400, the 111 Project under
Grant No. B20063, the National Natural Science Foundation
of China under Grant No. 12047501, and the Fundamental
Research Funds for the Central Universities.
\bibliography{DDD.bib}

\end{document}